\documentclass[aps,prd,twocolumn,floatfix,nofootinbib]{revtex4}   

\usepackage{amsmath}    
\usepackage{graphicx}   
\usepackage{amsfonts}

\begin{document}

\title{Derivation of Hamiltonian mechanics from determinism and reversibility}
\author{Gabriele Carcassi}
 \affiliation{Brookhaven National Laboratory, Upton, NY 11973}
 \email{carcassi@bnl.gov}
\date{November 9, 2012}

\begin{abstract}
We put forth the idea that Hamilton's equations coincide with deterministic and reversible evolution. We explore the idea from five different perspectives (mathematics, measurements, thermodynamics, information theory and state mapping) and we show how they in the end coincide. We concentrate on a single degree of freedom at first, then generalize. We also discuss possible philosophical reasons why the laws of physics can only describe such processes, even if others must exist.\end{abstract}

\maketitle

\section{Introduction}
Understanding is being able to look at something from different perspectives, and realizing it's the same thing. At least to me anyway. There are parts of physics that overlap with philosophy and math, and I find that a better understanding of the connection between them provides an insight that none of them can provide alone.

In this work we will look into fundamental questions such as: why does classical mechanics have this particular mathematical form? Could we have something different, or is it necessary? If so, what are the assumptions, either physical or philosophical? And why are they encoded in math in the way they are? What limitations do those assumptions bring?

The answer that I believe I have found is that classical mechanics, in the Hamiltonian formulation, is necessary if we are describing a deterministic and reversible process for an infinitesimally reducible homogeneous body, one that you can think of being made of an infinite amount of indistinguishable parts. \footnote{Each of these terms has multiple meanings in different disciplines, so I'll try my best to clarify the flavor that is needed here.} That is: classical mechanics can be derived from those assumptions.

Instead of taking a direct abstract approach, I'll first "reverse engineer" from the equations, adding meaning to each piece. It's less tedious and more rewarding as you will probably have familiarity with many of the pieces, and we can concentrate on the connections. I think it will also help to clarify exactly what I mean for each individual term. We will then restart "from scratch", defining the state space from first principles, and arrive to the same conclusions.

\section{The Mathematics}

First we should review the math. I'll frame it from a somewhat unusual angle, which can be later mapped to a direct physical meaning. Hamilton's equations are usually written as:

\begin{align*}
\frac{dx}{dt} &= \frac{\partial H}{\partial p} \\
\frac{dp}{dt} &= - \frac{\partial H}{\partial x}
\end{align*}
which I find confusing: on the left side you have functions of $t$ while on the right side you have functions of $x$ and $p$. You cannot substitute $x(t)$ and $p(t)$ in $H$, or you won't be able to take the derivatives. What we really mean by those equations is:

\begin{align*}
\frac{dx}{dt} &= \frac{\partial H}{\partial p} \bigg|_{x(t), p(t)} \\
\frac{dp}{dt} &= - \frac{\partial H}{\partial x} \bigg|_{x(t), p(t)}
\end{align*}

That is, we take the derivative and then calculate it at the given position and momentum. This equation is really telling us two things, and I'd like to break them up to make that clear. The first part is:
\begin{align*}
\frac{dx}{dt} &= S^x(x(t), p(t)) \\
\frac{dp}{dt} &= S^p(x(t), p(t)) \\
\frac{dP}{dt} &= \vec{S}(P(t))
\end{align*}
where $P$ is our point in phase space. This tells us that the evolution is continuous in $t$, and that it's a function of the current state of the system. That is, given the state, we are always going to move in a particular direction. $\vec{S}$ is the vector field that tells us where each point is moving, and the field lines are the trajectories they are going to follow in phase (state) space.

The second part is:
\begin{align*}
S^x &= \frac{\partial H}{\partial p} \\
S^p &= - \frac{\partial H}{\partial x} \\
\vec{S} &= \left(
                  \begin{array}{cc}
                    0 & 1 \\
                    -1 & 0 \\
                  \end{array}
                \right) \vec{grad}(H)
\end{align*}

This tells us that our $\vec{S}$ admits a potential $H$, and the relationship is the gradient rotated $90$ degrees. This means $\vec{S}$ is pointing in the direction of constant $H$, that the field lines of $\vec{S}$ are the lines of equal $H$. We also have the following relationships:
\begin{align*}
div (\vec{S}) &= \frac{\partial}{\partial x} S^x + \frac{\partial}{\partial p} S^p \\
              &= \frac{\partial}{\partial x} \frac{\partial H}{\partial p} - \frac{\partial}{\partial p} \frac{\partial H}{\partial x} \\
              &= curl(\vec{grad}(H)) = 0
\end{align*}
First of all, the divergence of $\vec{S}$ is the curl of the gradient of $H$. This is possible because of the $90$ degree rotation and because in two dimensions the curl is a scalar, like the divergence. Second, the field $\vec{S}$ is solenoidal, which means:
\begin{itemize}
  \item the net flux of $\vec{S}$ within a region is zero: the area of phase space that enters is equal to the area that exits
  \item any region transported by $\vec{S}$ will preserve its area
\end{itemize}
These are not just consequences: these are necessary and sufficient conditions. They are equivalent to requiring that the divergence is zero.

Once these conditions are required, we can construct the potential $H$. Suppose, in fact, that you do have a vector field for which the flux flowing into any region is zero. Consider two points on the boundary of a region: the flux from one side equals the flux on the other side. Now change just one of the half boundaries: the new region still has zero flux, so the flux from the new half boundary is equal to the flux of the previous one, and to any other we would construct. In other words: the flux across a line only depends on the two extremes. You can now define a potential $H$ by first arbitrarily choosing a point for which $H$ is zero, and letting $H$ at any other point be the flux across a line between the reference and this point. You can then show that this is indeed the potential $H$ we started from\footnote{In two dimensions this potential is sometimes referred to as the stream function.\cite{white} With multiple degrees of freedom, the Hamiltonian can't simply be the stream function or the vector potential of the flux.}. This is just a sketch of a demonstration, but should give you the idea. This is very similar to constructing a scalar potential for an irrotational field.\cite{divgradcurl}

To sum up, Hamiltonian mechanics is telling us two things: the change of state is continuous and a function of the state; the flux of the change is 0 or, equivalently, the area of phase space is conserved. What do these statements mean physically?

\section{Measurements}

The fact that the change of state is a function of the state itself should make us think of determinism: the next state is a function of and only of the previous state. If we start from that premise, and assume the evolution continuous in time, we already have:
\begin{align*}
P(t+dt) &= P(t) + \frac{dP}{dt} dt \\
P(t+dt) &= P(t) + \vec{S}(P(t)) dt \\
\end{align*}
We should realize, though, that this is not sufficient to have actual determinism. A measurement is not just a value, but an uncertainty as well. Suppose, for example, the uncertainty increased dramatically from the initial state to the final state. Yes, theoretically the future state is determined but in practice it is not. So we should call deterministic only a process for which the future state can be determined at least at the same level of accuracy. Conversely, we will call reversible a process for which the initial state can be reconstructed from the final state at least at the same level of accuracy. If the uncertainty decreases in time, then this is not possible.

To sum up: a process for which the uncertainty does not increase in time is deterministic; a process for which the uncertainty does not decrease in time is reversible; a process for which the uncertainty remains the same is both deterministic and reversible. And these are the processes we are interested in.

As a measure of uncertainty we will use the determinant of the covariant matrix:
\begin{align*}
|\Sigma| &= \left|
                  \begin{array}{cc}
                    \sigma^2_x & cov_{x,p} \\
                    cov_{p,x} & \sigma^2_p \\
                  \end{array}
                \right|
\end{align*}
and we set:
\begin{align*}
|\Sigma(t)| &= |\Sigma(t+dt)| \\
&= |J \Sigma(t) J^T| \\
&= |J| |\Sigma(t)| |J| \\
|J| &= \pm 1
\end{align*}
where $J$ is the Jacobian matrix of the transformation. We calculate the Jacobian by taking the coordinates of the transformed point $P(t+dt)$ and deriving them.
\begin{align*}
J &= \left(
                  \begin{array}{cc}
                  \frac{dx(t+dt)}{dx}  & \frac{dx(t+dt)}{dp} \\
                  \frac{dp(t+dt)}{dx}  & \frac{dp(t+dt)}{dp} \\
                  \end{array}
                \right) \\
  &= \left(
                  \begin{array}{cc}
                  1 + \frac{\partial}{\partial x} \frac{dx}{dt} dt  & \frac{\partial}{\partial p} \frac{dx}{dt} dt \\
                  \frac{\partial}{\partial x} \frac{dp}{dt} dt & 1 + \frac{\partial}{\partial p} \frac{dp}{dt} dt \\
                  \end{array}
                \right) \\
  &= \left(
                  \begin{array}{cc}
                  1 + \frac{\partial}{\partial x} S^x dt  & \frac{\partial}{\partial p} S^x dt \\
                  \frac{\partial}{\partial x} S^p dt & 1 + \frac{\partial}{\partial p} S^p dt \\
                  \end{array}
                \right) \\
 &= 1 + \frac{\partial}{\partial x} S^x dt +\frac{\partial}{\partial p} S^p dt = \pm 1
\end{align*}
\begin{align*}
\frac{\partial}{\partial x} S^x dt +\frac{\partial}{\partial p} S^p dt = 0
\end{align*}
Only the positive solution is possible under the infinitesimal transformation which leads to the second condition we needed: divergence of $\vec{S}$ is zero. Since the Jacobian represents how much the space is stretched at each point, requiring that the Jacobian is unitary everywhere means that phase space is not stretched overall: if you stretch it locally in one direction than you must do the opposite on the other. If we imagine a uniform distribution over a region of phase space, it should be intuitive to see that if the region stretched or shrunk it would lead to a different uncertainty.

We have seen that conservation of uncertainty means conservation of phase space and therefore Hamiltonian mechanics. There is something unsettling, though: we think of uncertainty as being in the eye of the beholder. We do not consider it a characteristic of the system, it's not part of its ontology. But this requirement affects the evolution, giving us the equation of motion. So let's reframe the same argument in terms of thermodynamics and energy, which we do consider a property of the system.

\section{Thermodynamics}

First of all, let's assume we have a system undergoing a deterministic and reversible change. The final state is known given the initial state, but it also means the transformation is continuous: to be reversible in the thermodynamic sense we will assume the process is quasi-static so the change has to be small. This gives us:
\begin{align*}
P(t+dt) &= P(t) + \frac{dP}{dt} dt \\
P(t+dt) &= P(t) + \vec{S}(P(t)) dt \\
\end{align*}
which is the first equation. For the second equation: if the system is truly reversible, it means that no energy is lost to heat; if the system is truly deterministic, meaning that the future evolution \emph{only} depends on its state, it must be isolated, or the state would have to include the one of the external system acting on the first.\footnote{Let's clarify this with an example. Consider a car going at a constant speed. At first glance, you may consider this system deterministic because you can seem to predict the position and the (constant) momentum. The truth is that the system is under two forces: the friction of the air and the push of the engine. At some point, the car is going to consume its gas and the speed will start decreasing. We then realize that our initial assumption that the system was determined only by its position and momentum was wrong: we needed at least to take into account the level of gas in the tank to predict when the car would slow down. If we add that to our state, we notice we do not have a reversible system, as the energy from the tank is being lost to friction.} Therefore, we need to assume our system does not acquire or lose energy to the outside, or to other degrees of freedom: it needs to conserve energy. If we were to apply a force on the system, we would not be able to extract or inject energy while returning it to its initial state. In other words: the energy absorbed by the system on a closed trajectory in phase space has to be zero. This means:
\begin{align*}
\Delta E &= \oint dE = 0 \\
         &= \oint (dE_{kinetic} + dE_{work}) \\
         &= \oint (v dp - F dx) \\
         &= \oint (\frac{dx}{dt} dp - \frac{dp}{dt} dx) \\
         &= \oint (S^x dp - S^p dx) \\
\end{align*}
The energy absorbed goes either into kinetic energy $v dp$ or into work performed against the system $- F dx$. We recognize $v$ and $F$ as $S^x$ and $S^p$. As you can see, the absorbed energy along a path is the flux entering through it; thus energy conservation leads to zero flux over any region of phase space. These are the same condition.

Note that if the system absorbed energy along a closed trajectory, the integral would be positive and we would have flow going inside of the enclosed region; since we are spending energy to maintain the closed trajectory, we can assume that energy is being dissipated in some other degree of freedom, so we will call this case non-reversible. This is also the case were our measurement uncertainty decreases. If the system released energy along a closed trajectory, the integral would be negative, we would have flow going outside of the region; since we are extracting energy from the closed trajectory, we have to assume that energy is being added from some other degree of freedom, external to the system, so we will call this case non-deterministic because the system is not isolated and its behavior is dependent on an external system and its state. Non-deterministic in this sense does \emph{not} imply the behavior of the system is random, but that it is determined by some \emph{other} system or degree of freedom. This is also the case where our measurement uncertainty increases. You may find these definitions odd, mainly because they use standard language in a slightly non-standard way. But you should see that this particular meaning holds for different perspectives. So, in this sense, they are more useful and more "true".

To recap, the flow within a region of phase space is actually the energy absorbed by the system along its boundary. We called non-reversible the case of positive energy absorbed, and negative flux (the flux flows out of the region); we called non-deterministic the case of negative energy absorbed, and positive flux. While we know what the flux corresponds to physically, we don't have a good idea of what the area in phase space is. We know it's related to uncertainty, but uncertainty is not a property of the system.

\section{Information}

Let's assume we have a distribution $\rho(x,p)$ instead of a point particle. This, actually, makes a lot more physical sense because macroscopic systems are distributed in space, and a point particle is simply a limit where that distribution is concentrated. Let's suppose, again, we have a deterministic and reversible evolution, where each of the elements of the distribution can be followed independently. That is, we assume the system is infinitesimally reducible: we can think it as made of an infinite amount of tiny elements, each with its own state and evolution. This implies a map between the initial points and final points in phase space, so that we can keep track of the evolution of each tiny element. If we also assume the evolution to be continuous in time, we have, like before:

\begin{align*}
P(t+dt) &= P(t) + \frac{dP}{dt} dt \\
P(t+dt) &= P(t) + \vec{S}(P(t)) dt \\
\end{align*}

To identify an element in the distribution, and track that specifically, we need to identify it from the others: we need some extra information. But if the evolution is truly deterministic and reversible, the extra information we need to identify an element must be always the same no matter at what time: once identified at any $t$ it is identified for all $t$.

That extra information is what is defined as information entropy. For a distribution over a discrete set it is given by:

\begin{align*}
I &= \sum \rho(i) log( \rho(i))
\end{align*}
where the $log$ is usually taken in base $2$. This represents the number of average bits of information required to identify one element within the discrete distribution. But, in our case, we have a continuous distribution in $x$ and $p$: clearly the information needed to identify an element is infinite; but it's also clear that it will be more difficult to identify an element from a continuous distribution from $0$ to $2$ meters than an element from $0$ to $1$ meter. Yes, mathematically they both require infinite information, but there \emph{is} a finite difference: the first one is twice as hard. In the continuous case, we define the continuous entropy:
\begin{align*}
I &= \int \rho(x, p) log( \rho(x, p)) dx dp
\end{align*}
which you can think of as the information needed to identify an element in the distribution minus the entropy for a uniform distribution from $0$ to $1$. That is: it's a relative number, it can be negative, and tells you the number of bits compared to that. So, for example, a distribution from $0$ to $2$ is twice as spread, so the continuous entropy will be $1$, because you need an extra bit of information. This quick explanation is not meant to derive these concepts, but just to give you an intuitive understanding so that we can link it with the rest of the discussion.

So, going back to deterministic and reversible processes: what we need is for this quantity, the continuous entropy, to be conserved. Under a generic transformation, the entropy transforms as:
\begin{align*}
I' &= I  + \int \rho(x, p) log(J) dx dp
\end{align*}
where $J$ is again the Jacobian of the transformation. For the entropy to be conserved no matter the initial distribution, we need to require that $|J|$ is one. Which is the same condition we found for measurements and leads to area conservation. But why do we need to conserve area to conserve the entropy? Each small element carries its own entropy contribution, and all of them are summed through the integral. If phase space is stretched, the distribution in that neighborhood is stretched too, so the value of the density $\rho$ at that point will decrease (or increase if compressed). Which will change the contribution to the total entropy for that element.

Intuitive arguments are easier to understand: if your distribution stretches, you will have to identify elements from a bigger range. If we imagine a continuous distribution over a region, the entropy increases with the area covered. In fact, $log(A)$ will give you the entropy: one unit of area gives you $0$ (because it's the reference), two units gives you $1$ (one bit more than the reference) and so on. So, conserving the area means conserving the informational entropy.

Now, if the entropy decreases, it's impossible to tell past states as we have less and less information about the system: this is the non-reversible case, the area shrinks, the flux is inward and, as we saw before, energy is lost by the system, uncertainty decreases. If the entropy increases, it's impossible to tell future states because we will need more and more information about the system: this is the non-deterministic case, the area grows, the flux is outward, energy is gained by the system, uncertainty increases.\footnote{I realize the explanations are very succinct: I am cramming many concepts into a few paragraphs. While they may take time to sink in, once they do you realize that the connection is that straightforward. And it's fascinating.}

We have a direct understanding of the flux (the energy) and of the area (the entropy which is related to uncertainty), but there is still one thing missing. When we define our bijective map in the first equation, we do not get the second equation for free. Both parts are driven by assuming deterministic and reversible processes but we seem to have to "double-dip" on the same assumption. Why is that? Can it be avoided?

\section{State space definition}

To answer those questions we are going to start from scratch and define states from the ground up. This has the extra benefit of clearly spelling out our assumptions and answers other questions, such as: why are states in classical mechanics vectors in a real space? Why are point particles so important to describe macroscopic objects (which are definitely not point particles)? This treatment is necessarily a bit abstract, and it will take considerations from math, physics and a bit of philosophy. I am not particularly an expert in all of them: the aim is simply to have enough ground work for the mathematician, the physicist and the philosopher to see where this is going. I believe that, to properly treat all the sides, one would end up with a much much longer work which nobody would be able to read. Details can kill you.

So, let's start from scratch: we have a system. This system can be in different configurations, which we call states. To each state we assign a label that identifies the state. A label can typically be related to:
\begin{itemize}
  \item the way it was prepared - it's the state I prepared with the following settings
  \item an ideal measurement at that time - it's the state that has this momentum and this position
  \item the future evolution - it's the state that will bend right if I put it in a magnetic field
\end{itemize}
Now, whenever you talk about measurement in physics, especially in quantum, people drag in all sorts of questions, both philosophical and methodological. I just want to be clear that, for this treatment, we are not interested in those. The \emph{only} things we care about are the labels. \emph{How} you get to those labels, or \emph{why} can we have those labels in the first place are not concerns of ours at this time. And just to be clear: this is not to trivialize that work. Quite the contrary.

I believe that finding the right labels is the actual problem in science! Progress in physics resulted by correctly identifying new labels: energy, entropy, momentum for photons, lepton number and so on. The point is that, once that work is done, once we do have recipes for preparing states, once we do have a way to recognize them, we can create our abstract mathematical set of states, a label for each. And once we have that set, it does not matter what it actually represents: just its mathematical properties. Two different systems that map to the same mathematical structure will have similar properties. That's the beauty of math.

So we have in principle defined our set of states. But we can't do much with it. So we'll make the following assumption: the system is infinitesimally reducible. That is, the state of the whole system is the state of its parts, and the state of each part is in turn equal to the state of its parts, and so on ad infinitum. This assumption, which I'll call the classical assumption, is similar to the old Greek philosophical idea of infinite divisibility, to which Democritus' atomic idea is opposed. But there is a subtle and important difference: nothing here is divided physically, it's the state that, mathematically, can be seen as composed of two parts.

To clarify what I mean by reducibility in this context, and how is it different from divisibility, I'll make a few examples. Consider a living cell: there exists a process (mitosis) that starts with one cells and divides it into two. But the state of a cell is \emph{not} equal to the state of two cells. A cell is divisible into two cells but it's not reducible to two cells. Consider a magnet: we can think of it as made of a north pole and a south pole, so the state of the magnet can be described by the state of its poles. But if we divide it, we do not get a north pole and a south pole. A magnet is reducible to two poles but it's not divisible into two poles.

Consider a photon: it can decay into an electron-positron pair. But you can't think of a photon as made of an electron and a positron. A photon is divisible into a pair but it's not reducible to a pair. Consider a proton: as far as we know it's made of quarks and gluons. But if we try to divide it, we don't get isolated quarks and gluons. The proton is reducible to quark and gluons but it's not divisible into them.

If the state is infinitesimally reducible, we can just consider all the possible states of the infinitesimal constituents. The state of our system will then be fully described by a distribution of its constituents among those. For example, if the parts can only be in two states, then the state of our system is identified by how much is in the first state and how much is in the second. If the parts are fully identified by position and momentum, then we need to keep track what fraction of the state has what position and momentum, that is $\rho(x,p)$.

In mathematical terms, the classical hypothesis of infinitesimal reducibility implies that the state space is a linear vector space. The basis of that space represents the possible states of our infinitesimal elements. Why is this so important? Because we are enormously reducing the problem! To describe the evolution we do not have to handle each state individually, but we can limit ourselves to describing the evolution of each infinitesimal element of the base of our space. Naturally we can't always make that assumption, but when we can it's a really powerful assumption!

This tells us why point particles are so special: they are the basis, the states for our infinitesimal element. But what's important here is that we have defined them with a recipe of subdivision. That is: they don't exist by themselves. What exists are the full states, the infinite sum over infinite components. We are \emph{not} taking point particles and grouping them together into finite systems, we are taking finite systems and breaking them up into particles. They should really be called "point-like pieces". Later we'll see why the distinction is not just academic.

Ok, we have our state space, which is actually a linear vector space. Now we want to study the properties of a deterministic and reversible evolution. Turns out that infinite dimensional spaces hide things, so we'll start discrete. Let's consider a finite dimensional space of rank $N$, which has $N$ basis elements. As we said before, we can simply concentrate on the evolution of the bases, so what we need is simply a map $s_f = m(s_i)$ that tells us the final state $s_f$ given the initial state $s_i$. If the evolution is deterministic and reversible, the map is bijective, an invertible one-to-one correspondence.

Consider now a subset of $M$ basis elements and its evolution. We will have the following:
\begin{itemize}
  \item the net flux of states across the set is zero: the number of states that enter the set is equal to the number of states that exit
  \item the transformation preserves the number of states in the set
\end{itemize}
For the first, the set remains the same and elements go in and out. And because of the one to one mapping, each element that comes in must push another out. For the second, the set changes, but because of the one to one mapping the number of states is the same. Note that these statements are identical to the ones about area in the math section. Intuitively, we have already found the result: in the limit the area will be the measure of the number of states. Mathematically, we need to make the limit right. And I only have a slightly convoluted way to do that.\footnote{Many ideas and elements of this proof where taken from \cite{classical_dynamics}.}

The way I understand it, the trick is that we need to restrict the argument to depend \emph{only} on elements of a finite region, so that it does not matter whether the space of our parameters is open (i.e. infinite). So, let me first rephrase the argument in the discrete case. Let's still consider a set of $M$ basis. If we apply the map, the states will "move" to other states. Some of these states will go outside the subset and some will stay inside. Since each state will have to go somewhere, we can say that $M = N_{gi} + N_{go}$: the number of states that go inside ($N_{gi}$) plus the number of states that go outside ($N_{go}$) is equal to the states we started with. We can also apply the map in reverse, and see where they are coming from. Some are coming from outside the subset and some come from inside. We can write $M = N_{ci} + N_{co}$: the number of states that come from inside ($N_{ci}$) plus the number of states that come from outside ($N_{co}$) is equal to the number of states in the subset.

Now, consider a mapping between elements that stay inside. That is, that maps two states that come from inside and go inside. Since the map is bijective, we can make a one-to-one correspondence between the two. That is the number of elements that come from inside must be equal to the number of elements that go inside: $N_{ci} = N_{gi}$. If that's the case, it's also true that that $N_{co} = N_{go}$: the number of elements that enter the set is equal to the number of elements that leave the set. In other words, we have divided the set of $M$ states into two subsets in two different ways, but the sizes of those subsets are the same.

These two equalities represent the properties as before, but now the mapping does not go out of the set of $M$ states: the ones that stay inside are mapped to each other, the ones that come from outside to the ones that go outside (don't care how, but they are in the same number). This way we have a chance to keep the mapping when we take the limit.

If we let just $N$ go to infinity, the discussion does not change: $M$ remains finite and so do all the other numbers. For the continuous limit, we need to do more work. Let's start by transporting our discrete basis to the $(x,p)$ plane, that is each state is labeled with an $x$ and a $p$: it can be identified by a point in phase space.

Consider a region $\mathbf{R}$ of our parameters: it will contain a number of elements $M$ whose labels fall within the area. Let's call each $\xi_k$, $k \epsilon \{1, 2, ..., M\}$.  We can partition that area so that each state is in one cell, each cell has one state and all the area is covered. We could imagine the area of the cell to signify the uncertainty on our label, but it's not necessary and anyway it will not matter once the limit is taken. We chose the regions to be not-overlapping to signify that they form a basis: none of them could be reduced into the sum of each other. Note that we are still in the previous case: under a bijective map the number of elements do not change. If we assumed each cell to be of the same size, this would would mean that the area would not change either. But we needn't assume that: let the area of each cell $\mathbf{V}_k$ be different. And we have:
\begin{align*}
\sum\limits_{k=1}^M \mathbf{V}_k = \mathbf{V}
\end{align*}
where $\mathbf{V}$ is the total area. We can define a discrete density function $D'(\xi_k)$
\begin{align*}
D'(\xi_k) = \frac{1}{M\mathbf{V}_k}
\end{align*}
so that:
\begin{align*}
\sum\limits_{k=1}^M D'(\xi_k) \mathbf{V}_k = 1
\end{align*}
This sum is finite, no matter the size of $M$, and it's a sum of a function multiplied by a volume. As before, we require a bijective map to be defined on the elements, and, as before, we know that the number of elements that start in the set is the same as the number of elements that end in the set. Summing over those two gives us two different sums:
\begin{align*}
\sum\limits_{k=1}^{N_{ci}} D'(\xi_k) \mathbf{V}_k = F_{ci} \\
\sum\limits_{k=1}^{N_{gi}} D'(\xi_k) \mathbf{V}_k = F_{gi}
\end{align*}
$F_{gi}$ represents the fraction of the region $\mathbf{R}$ covered by the states that remain inside the region, while $F_{ci}$ represents the fraction covered by states that are coming from the region. If we divided into cells of the same size, these numbers would already be equal, but at this point they are not. What is always equal are the number of points $N_{ci}$ and $N_{gi}$.

It's time to take the limit. Let's increase $M$, so the packing gets tighter, each area $\mathbf{V}_k$ gets smaller and the distance between all points decreases. We do it so that all products $M \mathbf{V}_k$ remain finite. $D'(\xi_k)$ approaches the normalized density distribution function $D(\xi)$, which is assumed to be continuous and finite. The first sum becomes the integral:
\begin{align*}
\int_{\mathbf{V}} D(\xi) v = 1
\end{align*}
where $v = dx dp$. The other two sums become:
\begin{align*}
\int_{\mathbf{R}_{ci}} D(\xi) dx dp = F_{ci}
\int_{\mathbf{R}_{gi}} D(\xi) dx dp = F_{gi}
\end{align*}
These integral are the same: the more we pack the region, the more the difference between the areas of the cells becomes negligible. In the limit, each cell will cover the same area, but since the number of cells is the same because of the bijective mapping, the result is the same. Which means: the fraction covered by the region is the same, therefore the area is the same. This also means that the area going through the boundary is zero, and if we have an infinitesimal transformation, the flux corresponds to the change in area, so it will be zero. You can now deduce that all areas are conserved too: start with a region, transform it into the end region and consider the union of the two; the regions that go inside and comes from inside coincide with the start and end regions, which means they will have the same area. Phew.

As it is sometimes the case, something that is evident intuitively becomes convoluted to demonstrate mathematically. Naturally there may be a better proof, though I would not be surprised if there weren't. The important bit here to understand is that a state is a region of phase space, not a point. When we take the limit, we are assuming we are getting better and better at labelling our state: that we can ideally break up our finite system in smaller and smaller pieces, with more refined labels. The region covered by each state becomes infinitesimal, but not zero. This means that for each region we will have an infinite number of states, but a region double the size will still contain double the number of states. In other words: yes they are both infinite, but the ratio between them is finite and well defined. This is the same effect we saw when talking about informational entropy: you still need an infinite amount of information to identify each element, but double the size of the distribution and you double the amount of information. It is the same effect precisely because the cause is the same: you are doubling the number of cases, you are doubling the size of the basis.

The other important thing is that we are defining the map on the discrete case and taking the limit while maintaining the map. Physically, this means that our deterministic and reversible process has to be defined at the same resolution. That is: you have to be able to determine the future behavior of the system with the same uncertainty. This is obvious once you think about it, but it's precisely what you miss if you don't have the area conservation. The last interesting bit is that the area conservation is guaranteed only in the limit. In fact, in the discrete case you could either have each state covering a different amount of phase space, or you could even leave that undefined. In other words: the area conservation is a consequence of the continuity of the parameters $x$ and $p$.

This pretty much concludes the main discussion: we have been able to derive the Hamilton equations for one degree of freedom starting from determinism and reversibility. This also allowed for a more in depth look at those concepts, and a better understanding of the connection between seemingly disparate things. There are still details I believe could be improved (why is energy connected to the state flow? why does the evolution need to be continuous? why is a spatial degree of freedom covered by just position and momentum?), but the foundations are there and I believe are solid.

\section{Generalization to multiple degrees of freedom}

Generalizing to three degrees of freedom means working with a six-dimensional space. This will make it impossible to visualize what happens. Still, with the knowledge we gained before, there is plenty we can understand.

\textbf{Notation.} First of all, let's introduce some useful notation.\footnote{The mathematician will forgive me if I use some Riemannian geometry notation for something that is really symplectic geometry. I find that this helps create an intuitive understanding of what's going on geometrically, and will later allow making an interesting parallel.} $x_i$ will represent the different directions in space, and $p_i$ the different components of momentum.\footnote{As a reminder, space and momentum are parameters which we use to divide our states, and phase space is really the space of the basis for the division. Studying the evolution of the basis means studying the evolution for all states.} Greek letters will represent all directions in space and momentum. We will use the standard tensor notation, and as an analog to the covariant component we define:
\begin{align*}
S_{x_i} \equiv - S^{p_i} \\
S_{p_i} \equiv S^{x_i}
\end{align*}
We'll see later why this is useful and may even make sense.

\textbf{Thermodynamics.} This approach works as before. Energy absorbed along all paths needs to be zero, except now paths are in six dimensions. We have:
\begin{align*}
\Delta E &= \oint (S^{x_i} dp_i - S^{p_i} dx_i) \\
 &= \oint (S_{x_i} dx_i + S_{p_i} dp_i) \\
 &= \oint S_{\alpha} d\alpha = 0 \\
\end{align*}
the new notation is indeed useful. Stokes theorem tell us this is equivalent to setting the curl of $S_{\alpha}$ to zero:
\begin{align*}
\partial_{\alpha} S_{\beta} - \partial_{\beta} S_{\alpha} &= 0 \\
\end{align*}
and allows a potential $H$:
\begin{align*}
S_{\alpha} &= \partial_{\alpha} H \\
\end{align*}
which we can rewrite as:
\begin{align*}
d_{t}x^i &= S^{x^i} = S_{p_i} = \partial_{p_i} H \\
d_{t}p^i &= S^{p^i} = - S_{x_i} = \partial_{x_i} H \\
\end{align*}
the Hamilton equations.\footnote{Note that force acting on the system is conservative. Energy conserved along all paths means also energy conserved in spatial paths only, which is the requirement for a conservative force.} This is a neat compact form, but we skipped a few questions. What are the $S^{\alpha}$ components actually doing? And why are the $S^{\alpha}$ related to the $S_{\alpha}$ in that way?

\textbf{State space.} For this approach, we will assume we have three independent degrees of freedom (d.o.f.). The relationships between $x_i$s and $p_i$s will have to be the same as in the single d.o.f.: if they weren't, one degree of freedom could "tell" there were others, and they wouldn't be independent. But what is the relationship across them? We may be tempted to just assume they are orthogonal, but orthogonality in physical space may or may not have anything to do with the orthogonality in phase space.

To understand this better, consider a small rectangle in phase space and its projection onto a d.o.f. If the projected area equals the area of the rectangle, given that the area on the d.o.f. is a measure of number of states, we can make a one to one correspondence between states on our d.o.f. and states in the rectangle. They are dependent to the point that they are the same degree of freedom. If the projected area is zero, there is no relationship between them. Identifying one point within the rectangle will tell us nothing more about our d.o.f. In other words, they are independent. In phase space orthogonality means independence. So, yes $x_i$ and $x_j$ are orthogonal; not because of the spatial relationship, but because they represent independent d.o.f.

Note that this definition is consistent with other views. For measurements, if we have some uncertainty on one d.o.f. we should not have uncertainty on the others: since uncertainty was equivalent to area, surfaces on independent d.o.f. must be perpendicular. For information theory, entropy along independent d.o.f. must also be independent which also requires orthogonality.

We can now define an inner product: given two vectors $\vec{V}$ and $\vec{W}$ we sum the vector products of the component in each degree of freedom. This gives us the sum of the projected areas made by the rectangle that has those vectors as sides. The metric is a measure of states, the generalization of the area for the single d.o.f.
\begin{align*}
\vec{V} \ast \vec{W} &= V^{\alpha} \omega_{\alpha, \beta} W^{\beta} \\
\omega_{\alpha, \beta} &= \left(
                      \begin{array}{cccccc}
                        0 & 1 & 0 & 0 & 0 & 0 \\
                        -1 & 0 & 0 & 0 & 0 & 0 \\
                        0 & 0 & 0 & 1 & 0 & 0 \\
                        0 & 0 & -1 & 0 & 0 & 0 \\
                        0 & 0 & 0 & 0 & 0 & 1 \\
                        0 & 0 & 0 & 0 & -1 & 0 \\
                      \end{array}
                    \right) \\
\end{align*}

What must happen under evolution? Each degree of freedom will conserve its area, as we have seen. Orthogonal directions remain orthogonal because independent degrees of freedom will need to remain independent. In other words: the inner product that we defined needs to be conserved. Under an infinitesimal transformation, we have:
\begin{align*}
V^{\alpha} \omega_{\alpha, \beta} W^{\beta} &= V'^{\alpha} \omega_{\alpha, \beta} W'^{\beta}  \\
&= (V^{\alpha} + \partial_{\gamma} S^{\alpha} dt V^{\gamma}) \omega_{\alpha, \beta} ( W^{\beta} + \partial_{\delta} S^{\beta} W^{\delta} dt) \\
&= V^{\alpha} \omega_{\alpha, \beta} W^{\beta} + (\partial_{\gamma} S^{\alpha} V^{\gamma} \omega_{\alpha, \beta} W^{\beta} \\
 &+ V^{\alpha} \omega_{\alpha, \beta} \partial_{\delta} S^{\beta} W^{\delta}) dt + O(dt^2) \\
\end{align*}
Consistently with what we did before, we define:
\begin{align*}
S_{\beta} \equiv S^{\alpha} \omega_{\alpha, \beta}
\end{align*}
and since $\omega$ is antisymmetric, we have:
\begin{align*}
S_{\alpha} = - S^{\beta} \omega_{\alpha, \beta}
\end{align*}
So we see that, by analogy, we can think of them as covariant components. We substitute, simplify and disregard the higher order in $dt$:
\begin{align*}
V^{\gamma} W^{\beta} \partial_{\gamma} S_{\beta} - V^{\alpha} W^{\delta} \partial_{\delta} S_{\alpha} = 0 \\
\end{align*}
Since this equation must hold for all vectors, it must hold for any pair of components independently:
\begin{align*}
V^{\alpha} W^{\beta} ( \partial_{\alpha} S_{\beta} - \partial_{\beta} S_{\alpha} ) &= 0 \\
\partial_{\alpha} S_{\beta} - \partial_{\beta} S_{\alpha} &= 0
\end{align*}
which is the same condition we had with the thermodynamic approach.

Ok, we found this nice mathematical result by defining some kind of metric. But how do things move? As we said, the transformation has to preserve two things: areas on all d.o.f., and orthogonality between independent ones. The flux itself may be hard to visualize. But we can imagine a small hypercube of phase space. This will have faces along independent degrees of freedom, and faces perpendicular to the independent degrees of freedom. The first set can change into any parallelogram provided it preserves the area; the second set must preserve orthogonality so it can change into any rectangle of any size. The hypercube can rotate, so that it may mix d.o.f.. The volume can't change: it's given by the product of the areas on the independent degrees of freedom, and since those remain the same, the volume remains the same.

Last note: in special relativity, we define the metric $\eta_{\alpha, \beta}$ for an inertial observer; we find the transformations that preserve that metric; we obtain the Poincar\'{e} group. That is the most general set of transformations that gives us inertial observers. Here we set a metric $\omega_{\alpha, \beta}$ that allows us to count states on independent degrees of freedom; deterministic and reversible processes must conserve that metric so we find the most general group with that property; we obtain Hamilton's equations. The two metrics are defined on different spaces, have different physical meaning, but the mathematical process is the same.

\section{On the untenability of classical mechanics}

What I want to discuss here is a somewhat philosophical aspect. With the advent of quantum mechanics, we all know that classical mechanics is experimentally not tenable: it works for macroscopic systems but not for microscopic. So, in that sense, it's not "correct". And we put that in quotes because one may argue that no theory is "correct". At any rate: we know that it does not match experimental data, but could it have been "correct"? Or is there some fundamental problem with it?

My feeling is that classical mechanics is untenable from a logical perspective.

We saw how Hamiltonian mechanics applies to systems that are isolated. Not only are they isolated, but they can be reduced into infinitesimal parts, so that each element is isolated. There is no external exchange of energy. There is no external exchange of entropy. But if that's the case, we can't really tell that the system exists in the first place: we can't interact with it, we can't detect it. It's existence is a philosophical question, not a scientific one. To be able to say something exists, it needs to be interacting with something else in the universe.

Can't this be fixed by adding an external force? In fact, isn't gravity always interacting with everything? The problem is that all these forces are "optional" and observer dependent. Optional because the evolution can be thought as free motion plus the effect of all forces, so you still allow for the possibility for completely isolated motion. Observer dependent because, with a suitable coordinate transformation, they can be made disappear.

But in practice we \emph{can} use it! So it must be good for something, right? If we can disregard that outside interaction, if the system is isolated "enough", if the external exchanges of energy cancel each other out, than the effect is negligible compared to other effects and our predictions work.

Ok, there seems to be some inconsistency if we are interested in deterministic and reversible processes. But why are they so important? Or, why do the basic laws we write have that property? Can't we just assume something else? I don't think so. First of all, we want to write laws that tell us how different events in the past have caused the present and how the present will cause events in the future. If I have this I will have that. Cause and effect already implies determinism. But the need for these types of processes is really linked to the way we label states to begin with. And this is where I go back and talk a bit about the problems I disregarded.

Suppose we have a set of states, and we label them with a property that has a different value for each state. Position, momentum, color, shape... Do not care what it is. Suppose I give you a way to determine that at time $t$ that particular property had a particular value. Suppose, though, that I don't give you a process to properly prepare it. In fact, no process in the whole universe can result in a state where that property is set to a precisely determined value. It's always prepared randomly. Now you have a problem: you measure that value, but that value is always the output of a random process. So your measurement is indistinguishable from a random generator. How can you be sure you are actually measuring something? So, for your measurement to make sense, you need a process that deterministically prepares the values.

Conversely: suppose I give you a way to prepare the value, but I don't give you a process to measure it. In fact, no process in the whole universe could tell that the initial state had a particular value. You can never reconstruct it. Now you have the opposite problem: you have a value, but nothing depends on it. How can you tell you actually prepared something? So, not only do you need a way to deterministically prepare the value, you need a way to tell it was there.

For our definition of state to be physically meaningful, the label needs to be part of a chain of at least one deterministic and reversible process. In principle, things could still exist that are not part of any deterministic chain. But they can't be the argument of physics: they are not states and you can't write equations of motions.\footnote{There is a slight difference in that the chain could be of states of different systems, while the assumption we used was that the system itself is deterministic and reversible. This would be an interesting aspect to explore further as it necessarily would lead to different equations.}

Another way to look at this is to focus on isolation. We can talk about an object, its position, momentum, color and so on, only because we can act on it independently from the rest. We talk about pianos, guitars and metronomes because we perceive them as distinct. In fact, reductionism relies on parts that can be studied independently. If we can't separate the world into objects we can't label anything. And the only way to be sure that something is isolated, is to show that, at least in some cases, its future state depends only on its current state: the piano will remain where I left it, the metronome will keep swinging at a predictable pace. To be able to talk about objects and their state we need isolation: we need determinism.

But wait a minute: what about random variables? We \emph{do} have statistical processes: we study those all the time and we write equations for them! Doesn't that contradict all I just said, thus proving it nonsense? No. For statistical processes, the label is a statistical property. Like the average or the standard deviation. You need a way to prepare a state with a given average, you need a way to measure the standard deviation. All that I said before, applies to those quantities as those are the labels. It does not apply to the elements of the distribution because those are not the labels. In other words: the labels stop at what can be \emph{known}, not at what things \emph{are}. Labels are not ontological.

So we have pinned down this paradox: if everything can be known, everything can be controlled, everything must be isolated; but if everything is isolated, it cannot be controlled or known. Under this light it may be conceptually reasonable why we need something like quantum mechanics, which I believe does solve partly this problem. Each quantum state has a deterministic part and a non-deterministic part. The first is the part that can be prepared, known, labelled and studied; the second part cannot be prepared, known, labelled or studied. The idea is that the universe is aware of the deterministic part through the continuous unavoidable interaction with the non-deterministic one.

\section{Possible generalization to quantum mechanics}

My hope is that this work can also be generalized to quantum mechanics with minor conceptual modifications. I have bits and pieces, and hopefully an overall picture, but in this business until you have most of the details you got nothing. My feeling is that the first part of the state space definition is the same, applying labels to states; while for the second part we will assume the system to be irreducible (i.e. we cannot trace the evolution of parts of a particle). This means we would need some amount of extra information (that we will never have) to fully define each part. Each state, then, should have some entropy associated with it, which means we cannot reduce the spread in position and momentum as we please. And you can see already familiar themes. My hope is that multiple solid arguments will require the state space to be a complex vector space. After that, many elements are easy to get: unitary evolution is very close to having the Schrodinger equation; hermitian operators for observables and the projection postulate are straightforward to derive from the scalar product. While I still have a lot of work ahead, the hope that the final derivation will be essentially similar for both classical and quantum is very appealing to me.

\section{Conclusion}

While there are clearly some details to polish, I think the broad picture is firm. It seems to fit like a jigsaw puzzle between concepts that appeared, at least to me, completely disconnected. None of the concepts are new per se, just their connection. And it is that connection that I think gives a deeper understanding.

I hope you enjoyed this work, and that it gave you new insights and ideas to chew on.


\begin{thebibliography}{5}
\bibitem{extend} Bibliography should be expanded, but haven't had the time.
\end{thebibliography}
\end{document}